\documentclass[]{iopart}

\usepackage{graphicx}        
\usepackage{psfrag}

\begin{document}
\title[Binary black hole evolutions with sixth order
       finite differencing]{Reducing phase error in long numerical binary 
       black hole evolutions with sixth order finite differencing}
\author{Sascha Husa, Jos\'e A. Gonz\'alez, Mark Hannam,
        Bernd Br\"ugmann, Ulrich Sperhake}

\address{Friedrich Schiller University Jena, Max-Wien-Platz 1, 07743 Jena, Germany}
\ead{sascha.husa@uni-jena.de}

\date{\today}

\begin{abstract}
We describe a modification of a fourth-order accurate ``moving
puncture'' evolution code, where by replacing spatial fourth-order
accurate differencing operators in the bulk of the grid by a specific
choice of sixth-order accurate stencils we gain significant
improvements in accuracy.  We illustrate the performance of the
modified algorithm with an equal-mass simulation covering nine orbits.
\end{abstract}


\section{Introduction}
\label{sec:intro}

In the last two years the numerical solution of the general relativistic
two-body problem has made a giant leap forward with a series of
breakthroughs in 2005 \cite{Pretorius:2005gq,Campanelli:2005dd,Baker05a}. More
than forty years after Hahn and Lindquist started the numerical investigation
of colliding black holes~\cite{Hahn64}, the field has now passed several
crucial milestones toward simulating general inspiral situations, such as
simulations of unequal-mass binaries and calculations of the gravitational
recoil effect and the evolution of black-hole binaries with
spin~\cite{Baker2006b,Gonzalez2006,Campanelli2006b,Campanelli2006c,Herrmann2007,Koppitz-etal-2007aa}. 
The latter have recently lead to the spectacular finding that
extremely large recoils are possible for spinning black holes 
\cite{Gonzalez:2007hi,Campanelli2007,Herrmann2007b}.

In order to fulfill numerical relativity's promise of providing useful
information to the gravitational wave data analysis community, it is desirable
to perform long numerical inspiral evolutions that allow us to cleanly match
fully general relativistic and 
post-Newtonian waveforms with error bars, and thus produce ``complete''
waveforms, which contain large numbers of gravitational-wave cycles from the
inspiral phase, as well as simulating the merger and ringdown phases.
Such simulations will be necessary for a sufficiently dense sample of the
black-hole binary configuration space. 

Comparisons with post-Newtonian results have already started, and several
groups have published results showing good agreement of various aspects of
non-spinning simulations with post-Newtonian predictions 
(see e.g.~\cite{Baker2006a,Baker2006c,Bruegmann:2006at,Buonanno06imr,Berti:2007fi,
Pan:2007nw,Ajith:2007qp}). 
Precise error estimates and detailed coverage
of the unequal-mass and spinning cases are however missing. One serious
technical problem is that performing such long evolutions with good accuracy
in the phase is still
computationally expensive --- at least for standard ``moving puncture''
finite-difference codes
\cite{Campanelli:2006gf,Baker:2006yw,Sperhake2006,Bruegmann:2006at,Herrmann2007b,Koppitz-etal-2007aa}.
In order to overcome phase inaccuracies in long evolutions, an alternative
route is provided by spectral methods, and significant progress has been made
in this direction by the Caltech-Cornell group
\cite{Scheel-etal-2006:dual-frame,Pfeiffer:2007}.

The initial data we evolve differ from the choice
in~\cite{Scheel:2006-NFNR-talk,Pfeiffer:2007}, and we cannot make a
direct comparison of the accuracy of the results since puncture
evolutions cannot start with excision initial data. It would certainly
be worthwhile in the future to conduct direct comparisons regarding
efficiency for different codes. We only quote some preliminary numbers
here to illustrate the fact that while ultimately the convergence
behavior of spectral codes is superior, for the purpose of
``gravitational wave astrophysics'', finite difference codes may still
be competitive: The performance of the Caltech/Cornell spectral code
has been quoted for long time medium resolution runs as roughly 10 (for a ``$64^3$''
grid configuration) and 27 (``$76^3$'' grid) CPU hours per $M$
\cite{Scheel:2006-NFNR-talk} (we will quote time and length units
of the total black-hole mass $M$, see \cite{Bruegmann:2006at}). The highest
resolution run presented here 
performs at 5.5 CPU hours per $M$ on an Intel Woodcrest 2.66 GHz dual core
processor. Both codes show satisfactory accuracy for long evolutions, and
we certainly expect both codes to undergo further optimizations. Long
evolutions that show fourth order convergence for most of the simulation have
also been presented by the Goddard group~\cite{Baker:2006kr}, using
impressively low spatial resolution, but no details are given on code
timing.

We have previously reported accurate evolutions for approximately two orbits
(initial separation of $D = 6.45 M$) in \cite{Bruegmann:2006at}, with an error
in the merger time of 0.2 \% 
at a computational cost of 505 CPU hours (1.44 CPU hours/M), 
and we have reported a merger time error of 0.5\% for $D = 8 M$
simulations \cite{HusaBostonLSC}. 
At larger initial separations the number of orbits is a steep function of the separation,
and phase accuracy rapidly decreases. Our fourth order code would thus require resolutions
which we find hard to tolerate for performing large parameter studies in the style
of \cite{Gonzalez2006}. 
In the context of finite differencing it is natural to consider higher
order methods. For example, it already turned out to be important
to move from second-order finite differencing to fourth-order finite
differencing as the feasible evolution time for puncture evolutions increased.

However, using higher order finite differencing for the types of codes used
to simulate
black-hole binary inspiral is not entirely trivial: for the moving-puncture
method, which is currently employed in the majority of the current codes, the
continuum equations become singular at the  location of the ``puncture'', and
one might worry about the robustness of finite-difference
schemes. Furthermore, current mesh refinement algorithms in the field are
based on the use of buffer zones, whose number depends on the stencil width of
the finite differencing scheme. In three spatial dimensions high-order
finite-difference schemes with wide stencils easily lead to a drastic decrease
in performance caused by the additional computational load due to the extra
buffer points. This gain in accuracy pertains in particular to the phase of
the evolution.

In the present paper we report on a first step to significantly improve the
accuracy of current finite-difference codes to evolve black-hole binaries
by using sixth order accurate finite differencing operators in the
bulk of the grid. We combine the sixth order accurate derivative operators
with fourth-order accurate dissipation operators [setting $r=3$ in
Eq.~(\ref{eq:dissipation})] and Runge-Kutta time integrators, and aggressively
reduce the number of AMR buffer zones compared to the number we would
theoretically require for sixth order convergence. The penalty in
computational cost is a rather moderate 30\% compared with our fourth-order code.
(We have compared the average speed over 100 $M$ of evolution time for our largest
grid configuration, which produces a 29\% increase in computational cost;
in our experience this is typical for our current code).

\section{Summary of the ``moving-puncture method'' 
as implemented in the BAM code}

There is a large freedom in writing the Einstein equations as a system of partial
differential equations, and much research has gone into finding optimal choices.
In this work we employ the currently most popular choice, the BSSN
system~\cite{Baumgarte99,Shibata95,Alcubierre02a,Gundlach:2005ta},
which so far is the only system for which results have been reported of
long-time black-hole binary simulations that do not rely on black hole
excision as has been used for example in
~\cite{Pretorius:2005gq,Scheel-etal-2006:dual-frame,Pfeiffer:2007}.  

We use the BSSN system together with the 1+log and gamma freezing coordinate
gauges~\cite{Alcubierre00a,Baker:2001nu,Alcubierre02a}
as described in \cite{Bruegmann:2006at} (choosing in particular the parameter $\eta$
in the gamma freezing shift condition as $\eta=2$ as we have done previously).
These gauge conditions allow the ``punctures'' to move across the grid (``moving puncture''
approach~\cite{Campanelli:2005dd,Baker05a}) and allow an effective softening of the
singularity in the metric associated with an internal asymptotic
region~\cite{Hannam:2006vv,Hannam:2006xw,Brown:2007tb}, which had been prohibited by the
traditional ``fixed punctures'' approach. 
The BSSN system is based on a conformal decomposition of the spatial geometry,
writing the physical spatial metric as $g_{ij} = \chi^{-1} \tilde
g_{ij}$ (following \cite{Campanelli:2005dd}). The blowup of the metric at the
``punctures'' is absorbed into the conformal factor $\chi$, which vanishes at
the ``puncture''. 

For our numerical evolutions we use the
BAM~\cite{Bruegmann97,Bruegmann:2003aw} code, 
which is designed to solve partial differential equations on structured
meshes, in particular a coupled system of (typically hyperbolic) evolution
equations and elliptic equations.  The complexity of the equations is
addressed by using a Mathematica package integrated into the code, which
produces C-code from Mathematica expressions in tensor notation. Using such a
system as we do in BAM, or as has been discussed in detail for the Cactus
environment in \cite{Husa:2004ip} drastically simplifies the modification of 
complex codes for black-hole binary simulations, as was required to adapt
codes from the  ``fixed puncture'' to the ``moving puncture'' paradigm, or in
the present case to implement the improved numerical algorithms discussed here.
The structure of the BAM code has also made it straightforward to
implement higher order finite differencing methods 
The computational domain is decomposed into rectangular boxes, following standard
domain-decomposition algorithms, and is parallelized with MPI~\cite{Gropp:1996:HPI}. Our
mesh refinement algorithm is based on the standard Berger-Oliger algorithm,
but with additional buffer zones, along the lines
of~\cite{Schnetter-etal-03b,Lehner:2005vc} as described in
\cite{Bruegmann:2006at} and summarized in the next section.
We essentially use a fixed-mesh-refinement strategy, with inner level
refinement boxes following the motion of the black holes. Typically we use
about 10 refinement levels (refining the grid spacing by factors of 2),
roughly half of which follow the movement of the black holes.

In order to represent black holes in the initial data, we use the so-called 
``puncture method'' \cite{Brandt97b}.
For these data it is well understood how to write the constraint equations in
a form suitable for numerical solution~\cite{Beig94,Dain01a}. Following the
approach of~\cite{Brandt97b} our initial data sets are chosen to be
conformally flat with Bowen-York extrinsic curvature. The momentum parameter
in the Bowen-York extrinsic curvature is determined from a quasi-equilibrium
condition at third post-Newtonian order as described
in~\cite{Bruegmann:2006at}. The elliptic constraint 
equations are solved in BAM with the pseudo-spectral collocation
code described in~\cite{Ansorg:2004ds}. 
AMR data are then obtained by barycentric interpolation, typically with
eighth-order polynomials for both the fourth- and sixth-order finite
differencing methods. The efficiency of the spectral solver is
sufficient to solve the initial data problem on a single processor.

\section{Sixth order finite differencing}

\subsection{Mesh refinement in the BAM code}

Our numerical evolution algorithm is based on a method-of-lines approach using
finite differencing in space and explicit fourth-order accurate Runge-Kutta
time stepping (with a fixed time step). 
We apply sixth-order accurate polynomial interpolation in space
between different refinement levels so that all spatial operations of
the AMR method (i.e.\ restriction and prolongation) are sixth-order
accurate, such that the second derivatives of interpolated values are
at least fourth-order accurate.  Although the time stepping used for
evolution is also fourth-order accurate through the Runge-Kutta
integrator, there arises the additional issue of how to provide
boundary values for the intermediate time-levels of the Berger-Oliger
algorithm that are not aligned in time with a coarser level (otherwise
spatial interpolation can be used). Using higher than third-order
interpolation has lead to spurious noise at mesh refinement boundaries
as described in~\cite{Bruegmann:2006at}. We therefore use third-order
interpolation in time, which introduces a second-order error within
the Berger-Oliger time-stepping scheme~\cite{Bruegmann:2006at}, which
however is not noticeable in typical runs as we have checked by running
with uniform (as opposed to Berger-Oliger non-uniform)
time-stepping. In summary, if the outer boundary is placed
sufficiently far away and if time-interpolation errors at refinement
boundaries are small, then fourth-order convergence can be observed.

A relatively straightforward modification of the standard  Berger-Oliger
scheme is to replace the single-point refinement boundary by a buffer zone
consisting of several points,
e.g.,~\cite{Schnetter-etal-03b,Lehner:2005vc,Csizmadia:2006rz}. For a
sufficiently large number of buffer zones (the product of number of points in
the stencil toward the mesh refinement boundary and the number of source
evaluations during a full time step of the coarse grid), 
no time interpolation is required and excellent results have been reported for 
this scheme~\cite{Lehner:2005vc} (note that special methods like~\cite{Baker:2005xe} 
seem to achieve similar performance).
For example, our fourth-order Runge Kutta scheme requires four source
evaluations, and if the lop-sided stencil with three points in one direction,
\begin{equation}
\label{eq:lopsided_4th} 
f'(x) = \frac{-3 f_{-1} -10 f_0 + 18 f_1-6 f_2+f_3}{12 h} - \frac{1}{20} f^{(5)}(x) h^4
         + O(h^5)
\end{equation}
is used, then the numerical 
domain of dependence for a given point has a radius of 12
points. Here and in the following we use the notation $f_j = f(x + j\Delta x)$. 
Therefore, it is possible to provide 12 buffer points at the
refinement boundary and to perform one RK4 time step with size three
stencils that does not require any boundary updates. Only after the
time step is completed, the buffer zones have to be repopulated. In
the context of Berger-Oliger AMR, the buffer update is based on
interpolation from the coarser levels. Since every second time step at
level $l$ coincides in time with level $l-1$, one can provide 24
buffer points, perform two time steps, and then update the buffer by
interpolation in space. With 12 buffer points, one can interpolate in
time to obtain data for the buffer points at intermediate time levels.

To use fewer than 12 buffer points, we can interpolate into all buffer
points before starting an RK4 update as described, and then evolve all
points except the outermost points located exactly on the boundary,
which are kept fixed at their initial interpolated value. The inner
points next to the boundary are updated using second-order finite
differencing for the centered derivatives and shifted advection
stencils for the advection derivatives.
Even though for large grids the number of buffer zones becomes
negligible, for the grid sizes that we use, the buffer points affect
the size of the grids significantly. For example, even for our largest
inner box size of 80 points in one direction, adding six points on
both sides instead of 12 or 24 points leads to 92, 104, and 128
points, respectively, which corresponds to a significant saving in the
number of points in 3d since $104^3/92^3\approx1.44$ and
$128^3/92^3\approx2.69$.
For clarity, we always quote grid sizes without buffer points,
because this is the number of points owned by a particular grid.
Experimentally we have found that for the fourth-order case using just six
buffer points leads to very small differences compared to 12 buffer points,
but even smaller buffer zones lead to noticeable differences. For simulations
with fourth-order accurate derivative operators, we have therefore chosen a
standard setup of RK4 with dissipation and lop-sided advection stencils, 6
buffer points, quadratic interpolation in time, and Berger-Oliger
time-stepping on all but the outermost grids following \cite{Bruegmann:2006at}.

As is common in numerical relativity, we use symmetric finite-difference
stencils for all spatial derivatives but the advection terms associated with
the shift vector, where we use lop-sided upwind stencils, see e.g.\
\cite{Zlochower2005:fourth-order} for the fourth-order accurate case.

\subsection{Artificial Dissipation}

In finite-difference codes targeted at smooth solutions of nonlinear
hyperbolic equations, it is common practice to add artificial dissipation
terms to all right-hand-sides of the time evolution equations, schematically
written as 
\begin{equation}
\partial_t {\bf u} \rightarrow \partial_t {\bf u} + Q {\bf u}.
\end{equation}
Such dissipation terms are very efficient at suppressing very high-frequency
waves, which are not part of the physical solution. This may be necessary for
numerical stability  \cite{Kreiss73}, but also to reduce numerical noise
generated at mesh-refinement boundaries. As has become rather common in
numerical relativity, we follow \cite{Kreiss73} and choose an operator ($Q$)
of order $2r$ as 
\begin{equation}\label{eq:dissipation}
Q = \sigma (-h)^{2 r - 1} (D_+)^{r} (D_-)^{r}/2^{2r},
\end{equation}
for consistency with a $2r -2$ accurate scheme, with $\sigma$ a parameter
regulating the strength of the dissipation.
As we have done in the past with our fourth-order code we choose the factor $\sigma$
as $\sigma = 0.1$ in the inner levels and  $\sigma = 0.5$ in the outer levels
(where the waves are extracted).

For high orders, these dissipation
stencils become rather large (seven points for the fourth-order case and nine
points for the sixth-order case). We therefore do not add dissipation terms
where these stencils would ``cross'' mesh refinement boundaries. Also, adding
dissipation terms with large stencils can lead to a loss of performance.
We have therefore attempted to combine the use of sixth-order accurate
stencils for the derivative operators with a fourth-order accurate dissipation
operator and time integrator and second-order time interpolation at mesh
refinement boundaries with an aggressively small number (6) of buffer points.

\subsection{Sixth order accurate finite difference operators}

For the sixth-order case we find that several choices for the
advection term stencils yield stable evolutions, but the lop-sided upwind
stencil which is closest to the symmetric case yields (probably not
surprisingly) by far the best accuracy, i.e.\ we use 
\begin{equation}
\fl
f'(x) =
   \frac{2 f_{-2}- 24 f_{-1}-35 f_0 + 80 f_1 - 30 f_2 +8 f_3 -f_4}{60 h}
  +\frac{1}{105} \frac{d^7 f(x)}{dx^7} h^6 + O(h^7).
\end{equation}
Alternative asymmetric choices would be 
\begin{eqnarray*}
\fl
f'(x) &=&
  \frac{-10 f_{-1} -77 f_0 + 150 f_1 - 100 f_2 + 50 f_3 -15 f_4 +2 f_5 }{60 h}
 - \frac{1}{42} \frac{d^7 f(x)}{dx^7} h^6 + O(h^7),\\
\fl
f'(x) &=& \frac{-147 f_0+360 f_1-450 f_2+400 f_3 -225 f_4 +72 f_5-10 f_6)}{60 h} 
 +\frac{1}{7} \frac{d^7 f(x)}{dx^7} h^6  + O(h^7),
\end{eqnarray*}
for the stencils that deviate more from the symmetric choice. 
We can see that the first choice has the smallest leading error term.
The symmetric stencil has an even smaller error term,
$$
f'(x) = \frac{-f_3 + f_{-3} - 9 (f_2-f_{-2}) + 45 ( f_1- f_{-1} )}{60 h} - \frac{1}{140} \frac{d^7 f(x)}{dx^7} h^6 + O(h^8),
$$
but does not show equally robust results, as is common for solving advection 
equations. For non-advection derivative terms we again use the standard
symmetric stencil, similarly for second derivatives in one direction
we use the symmetric stencil 
$$
f''(x) = \frac{-490 f_0 + 270 (f_1 + f_{-1}) - 27 (f_2 + f_{-2}) +2 (f_3+f_{-3}) ) }{180 h^2}
-\frac{1}{560} f^{(8)}(x) h^6 + O(h^8).
$$
For mixed derivatives, we use the stencils which result from a product of the symmetric
sixth order accurate first derivative operators.

\section{Results for long equal-mass evolutions}

\begin{table}
\caption{\label{tab:orbit_grids_BAM}
  Grid setups used for convergence test simulations. The notation in the ``Run''
  column is the same as we have used in~\cite{Bruegmann:2006at}. The quantities
  $h_{min}$ and $h_{max}$ (rounded to three digits) denote the finest and
  coarsest grid spacings, and $r_{max}$ is the location of the outer boundary
  (rounded to 4 digits), and all are in units of $M$. Also specified are the 
  numbers of processors used, maximal memory requirement in GByte (to be
  precise, we quote the resident size of the program, i.e.,~the physical
  memory a task has used), and average speed in $M$/hour 
  for the Kepler cluster at the University of Jena (using Intel dual Woodcrest
  CPUs running at 2.66 GHz). The number in bold are used to indicate
  individual simulations throughout this paper.} 
\begin{tabular}{@{}l|r|r|r|r|r|r}
\br
Run & $h_{min}$ & $h_{max}$ & $r_{max}$ & procs. &  mem. (GByte)  & M/hour  \\
\mr
$\chi_{\eta=2}[5\times {\bf 48}:5\times  96:6] $ & $1/32.0$ &  16    & 776.0 & 8  & 12.2 & 15.6 \\
$\chi_{\eta=2}[5\times {\bf 56}:5\times 112:6] $ & $1/37.3$ &  96/7  & 774.9 & 12 & 18.2 & 11.6 \\
$\chi_{\eta=2}[5\times {\bf 64}:5\times 128:6] $ & $1/42.7$ &  12    & 774.0 & 12 & 22.5 & 8.4  \\
$\chi_{\eta=2}[5\times {\bf 72}:5\times 144:6] $ & $1/48.0$ &  32/3  & 773.3 & 16 & 31.3 & 3.5  \\
$\chi_{\eta=2}[5\times {\bf 80}:5\times 160:6] $ & $1/53.3$ &  48/5  & 772.8 & 24 & 45.4 & 4.4  \\
\br
\end{tabular}
\end{table}
\begin{figure}[t]
\centering
\includegraphics[width=6cm]{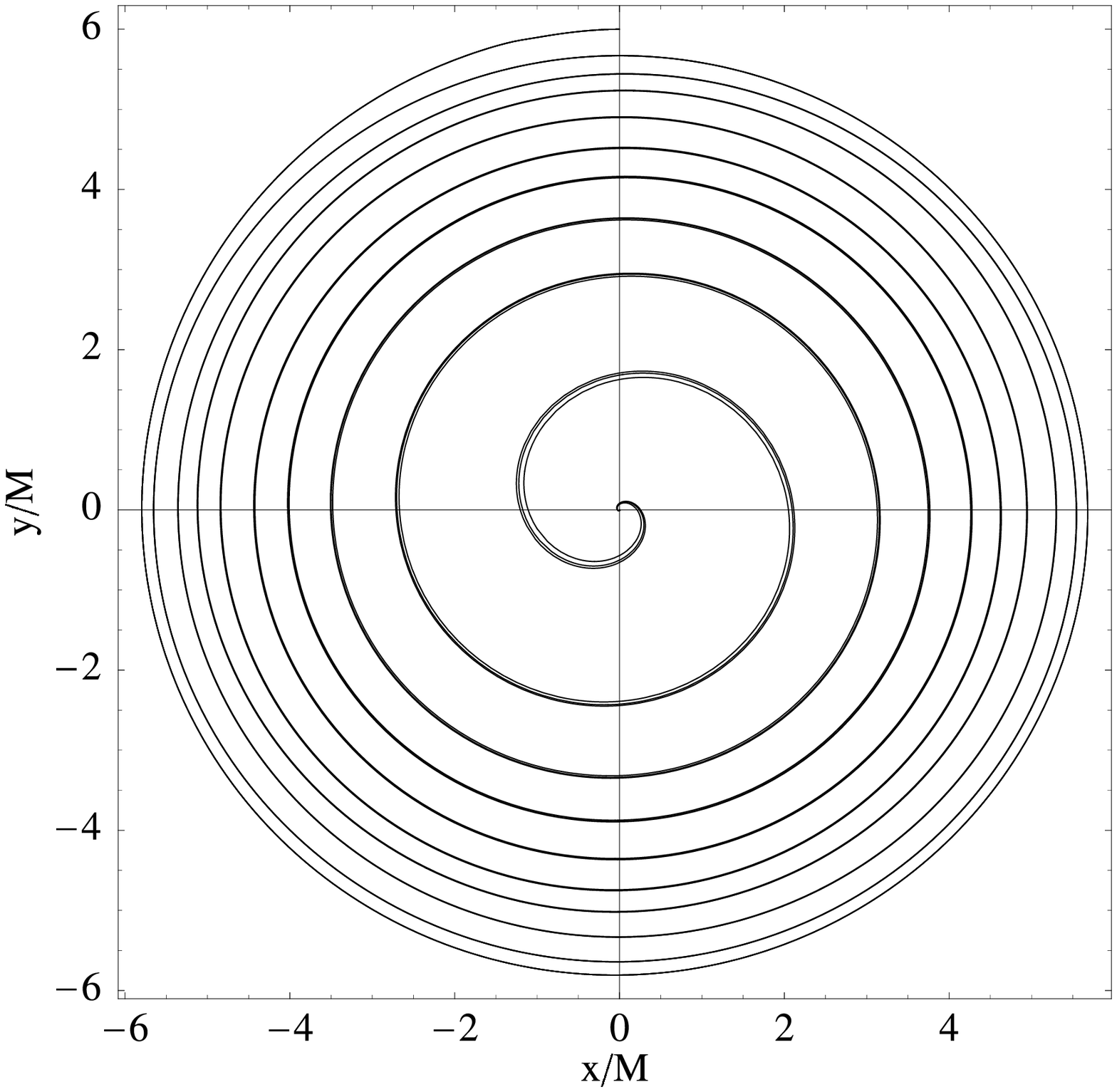}
\includegraphics[width=6cm]{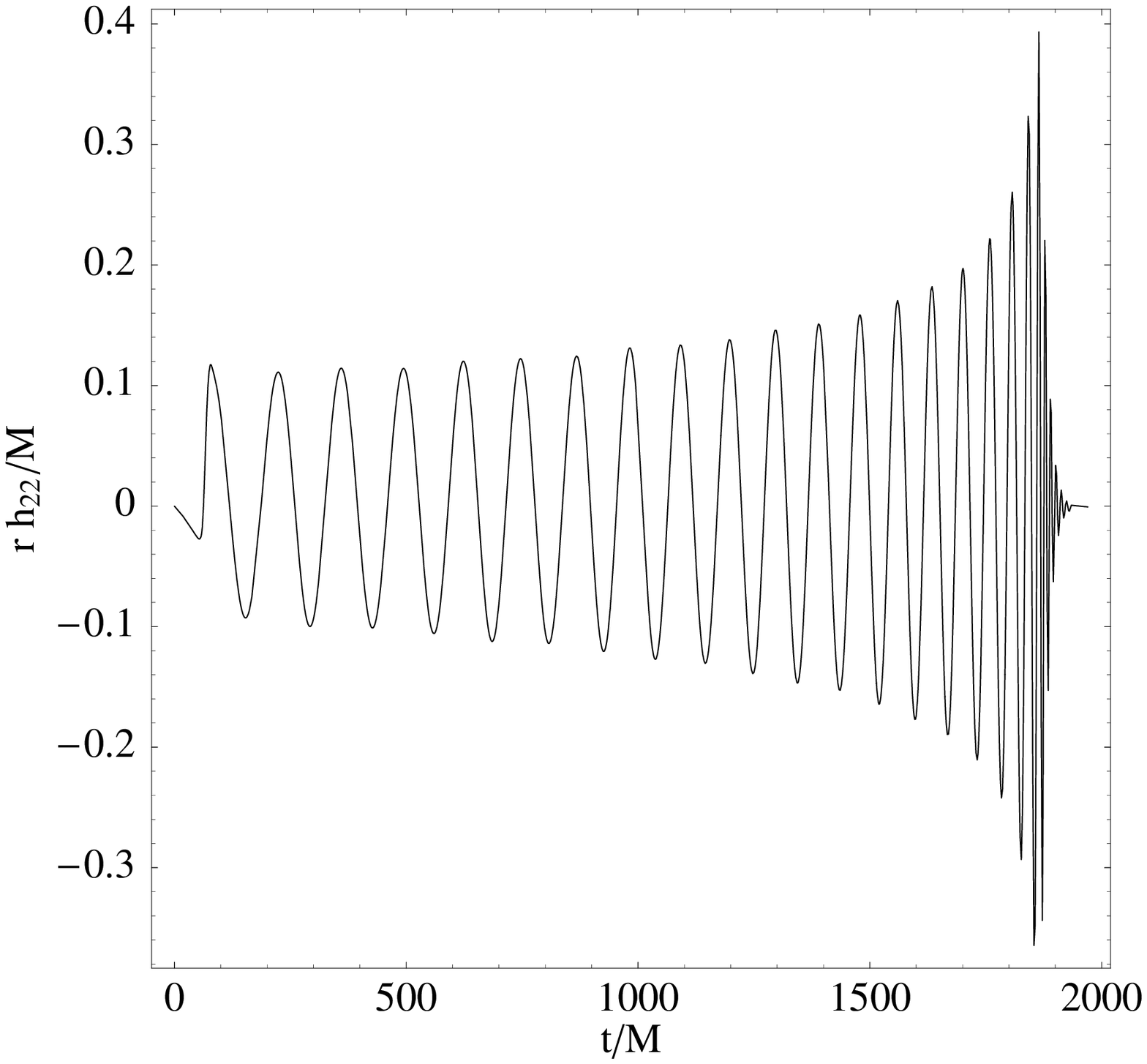}
\caption{Left panel: Coordinate tracks of the puncture location of one black hole
for simulations $\{64,72,80\}$. Only in the last few orbits are differences
between the three runs discernable. Right panel: the waveform plotted as the
real part of $r h_{22}$, as defined in 
\cite{Ajith:2007qp}.
}
\label{fig:tracks_strain}
\end{figure}
\begin{figure}[t]
\centering
\includegraphics[height=4cm]{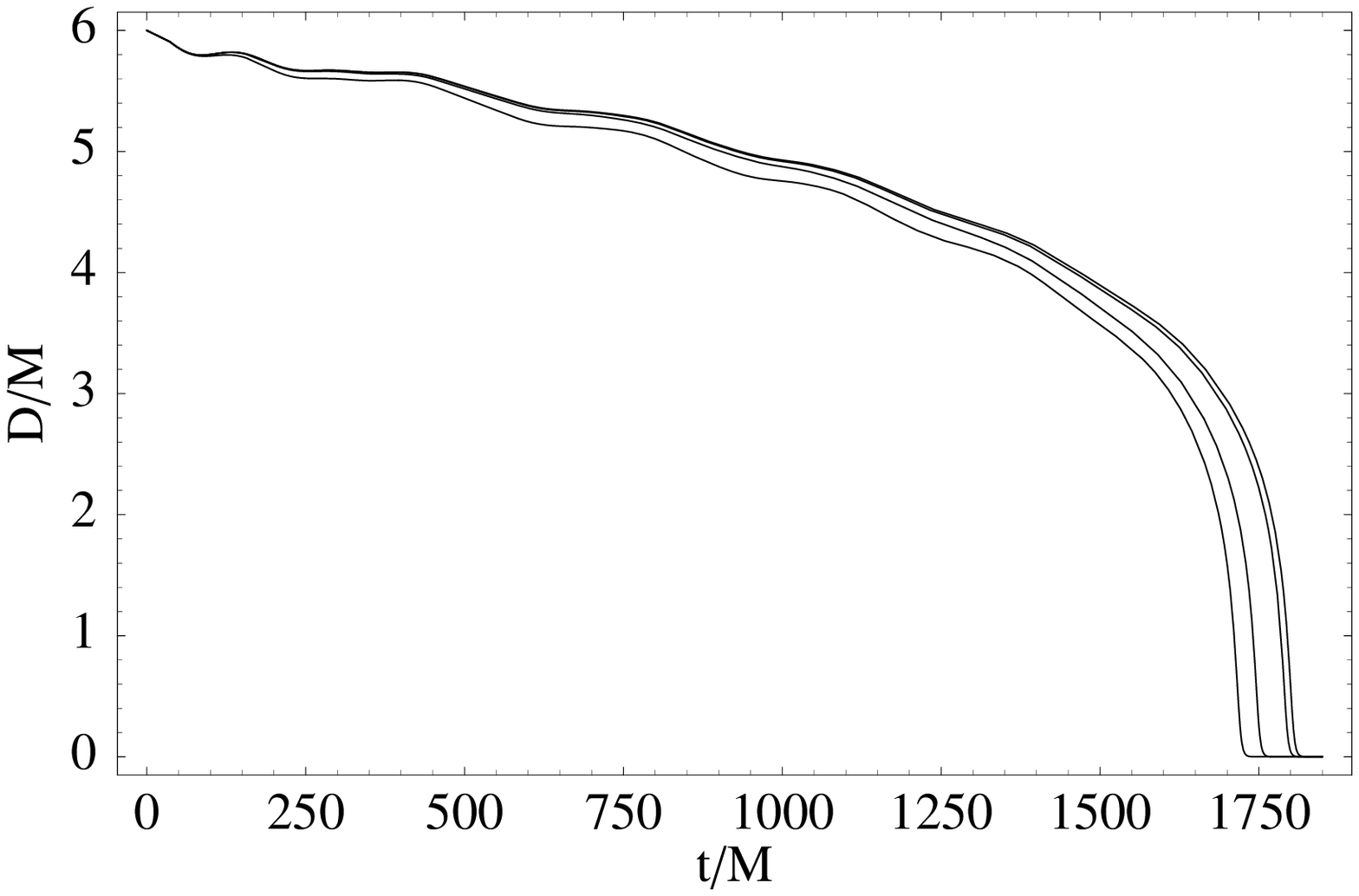}
\includegraphics[height=4cm]{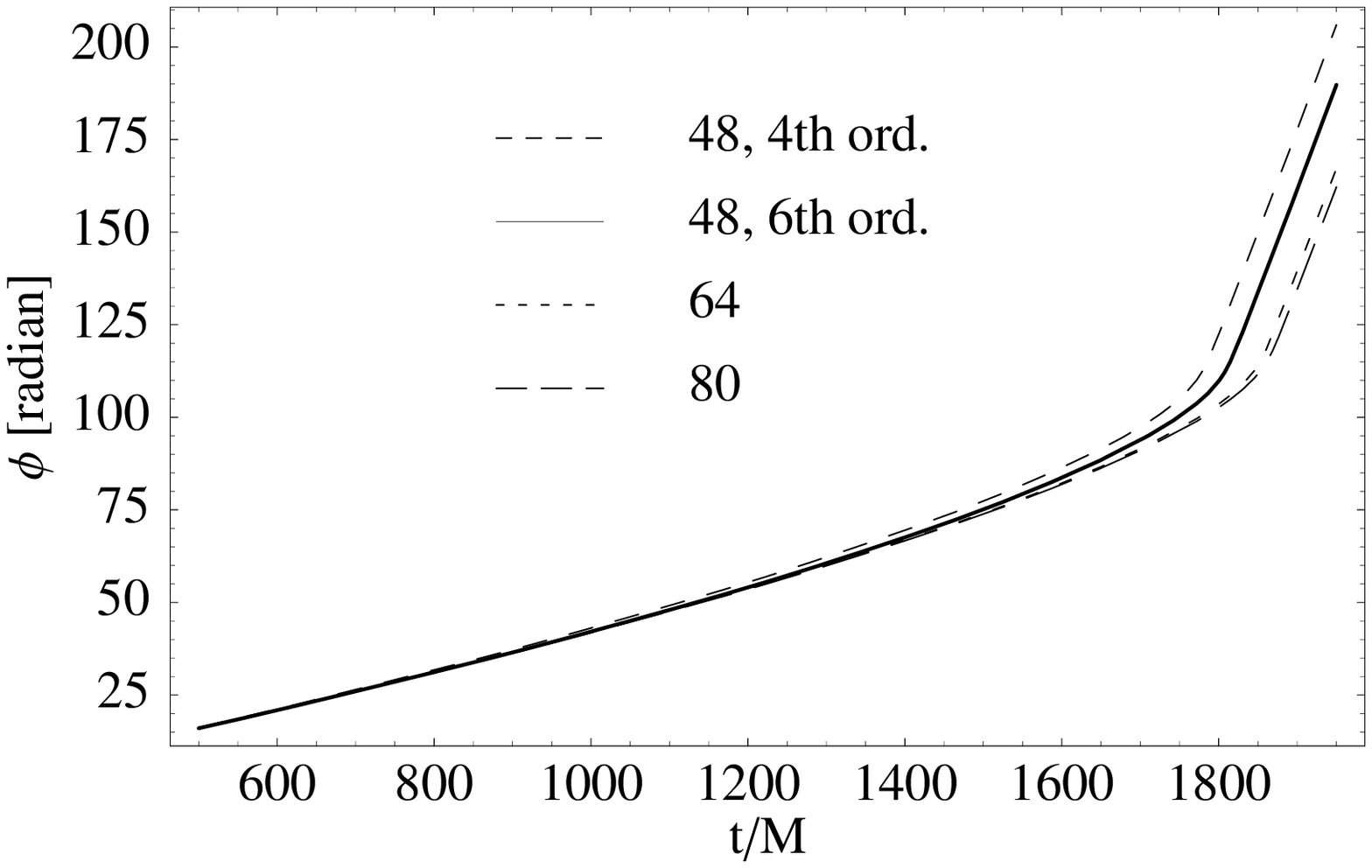}
\caption{Left panel: coordinate distance of the black holes for the fourth order version of the
$48$-configuration and sixth-order simulations $\{48,64,80\}$
in the order of increasing merger time. 
Right panel: the gravitational wave phase for the same runs.
The $72$ simulation would not be distinguishable from the $80$ simulation on the scale shown here.
}
\label{fig:distance_phase}
\end{figure}
\begin{figure}[t]
\centering
\includegraphics[height=4cm]{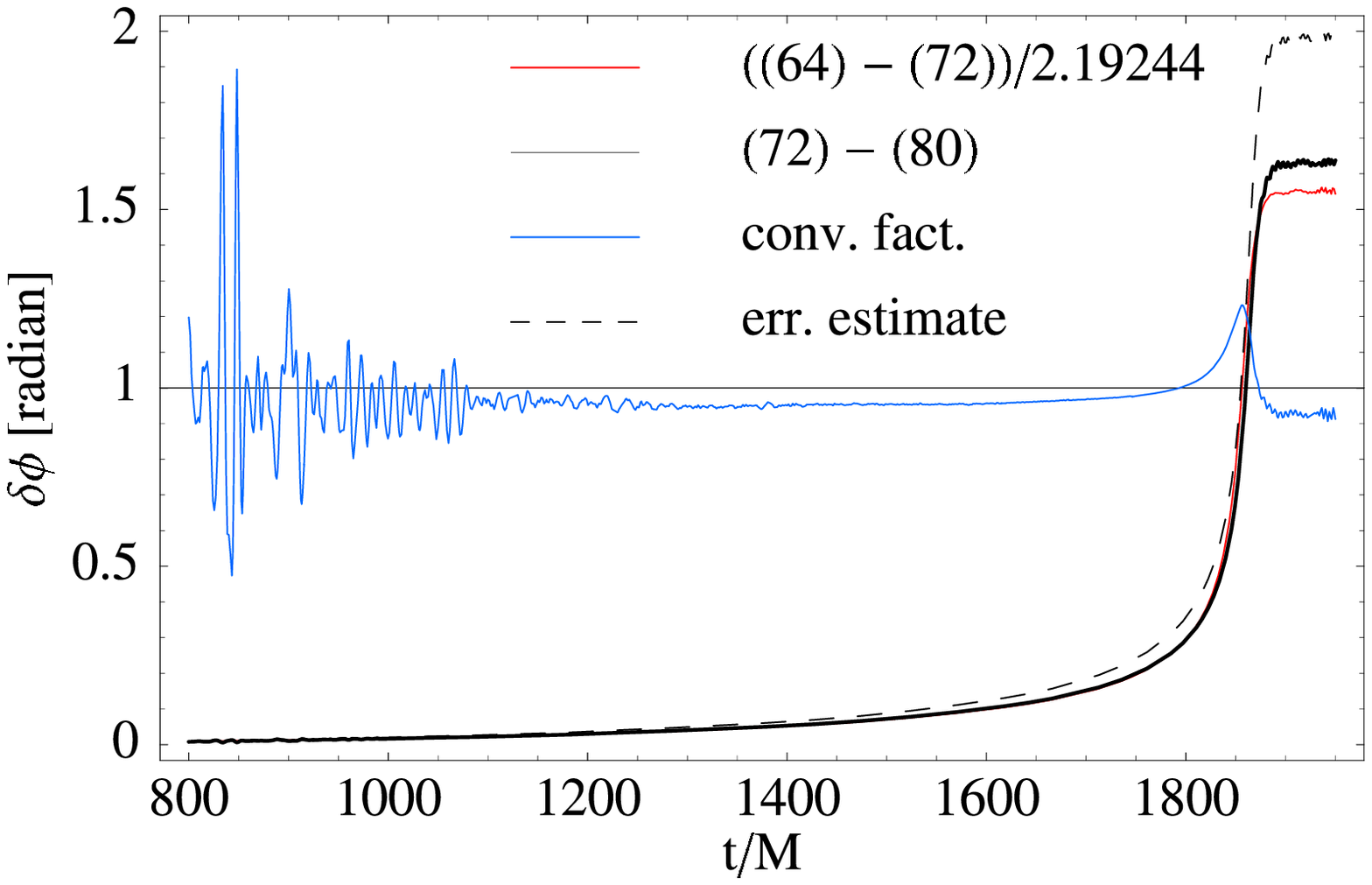}
\includegraphics[height=4cm]{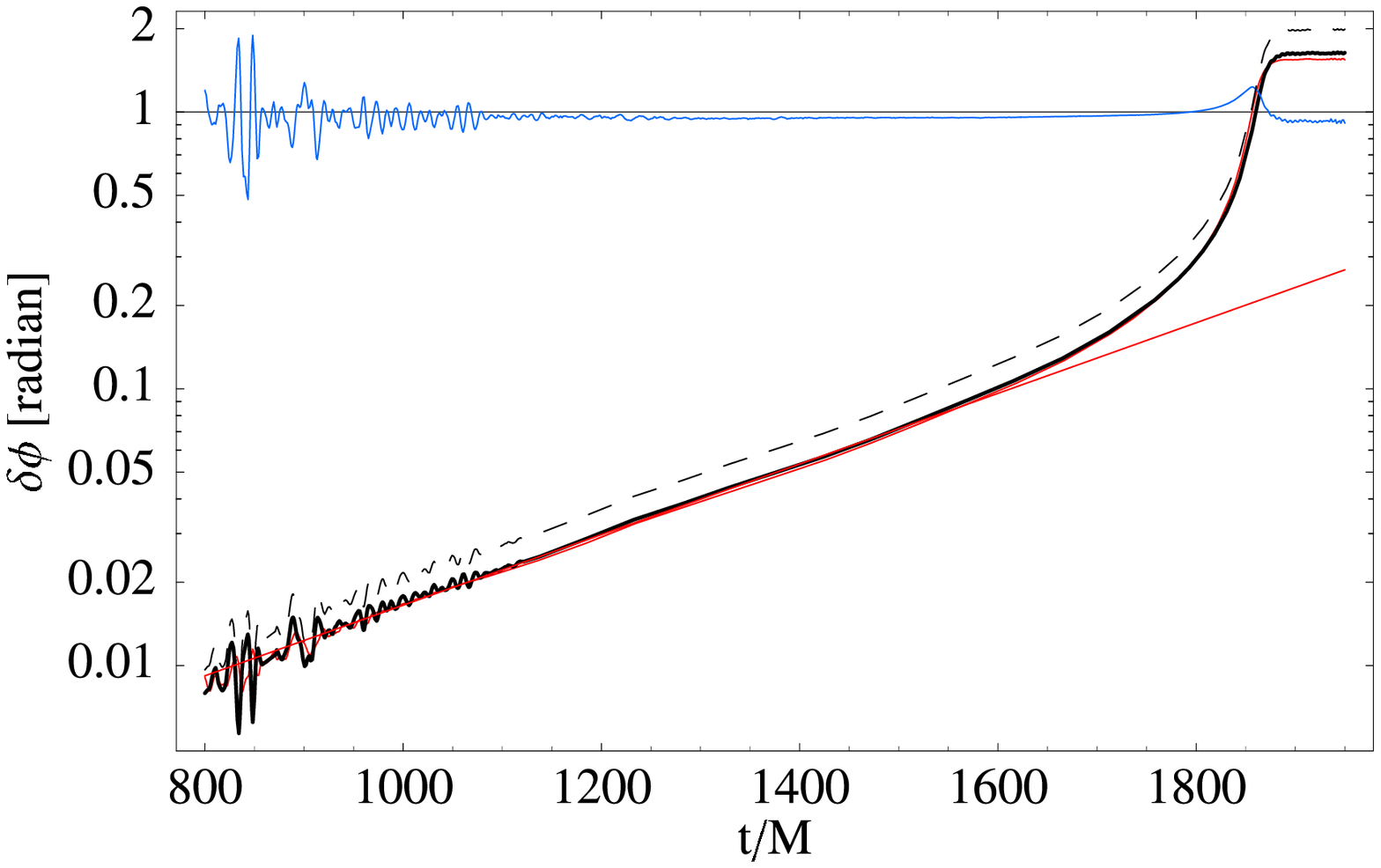}
\caption{Convergence test for the gravitational wave phase. Plotted are the 
difference between the $72$ and $80$ runs, and the difference between the $64$ and $72$ runs
rescaled for sixth-order convergence. Also shown is the convergence factor divided by $6$,
which shows a ``glitch'' around the time that the phase increases very
sharply, and the error estimate after performing Richardson extrapolation. 	
The left panel shows a linear scaling, the right panel shows the same plot with a logarithmic scaling
to emphasize the slow but clean exponential growth $\delta \varphi = 0.0117
\exp{ 0.003 t/M}$ of the phase error at intermediate times. 
}
\label{fig:wave_conv}
\end{figure}
\begin{figure}[t]
\centering
\includegraphics[height=4cm]{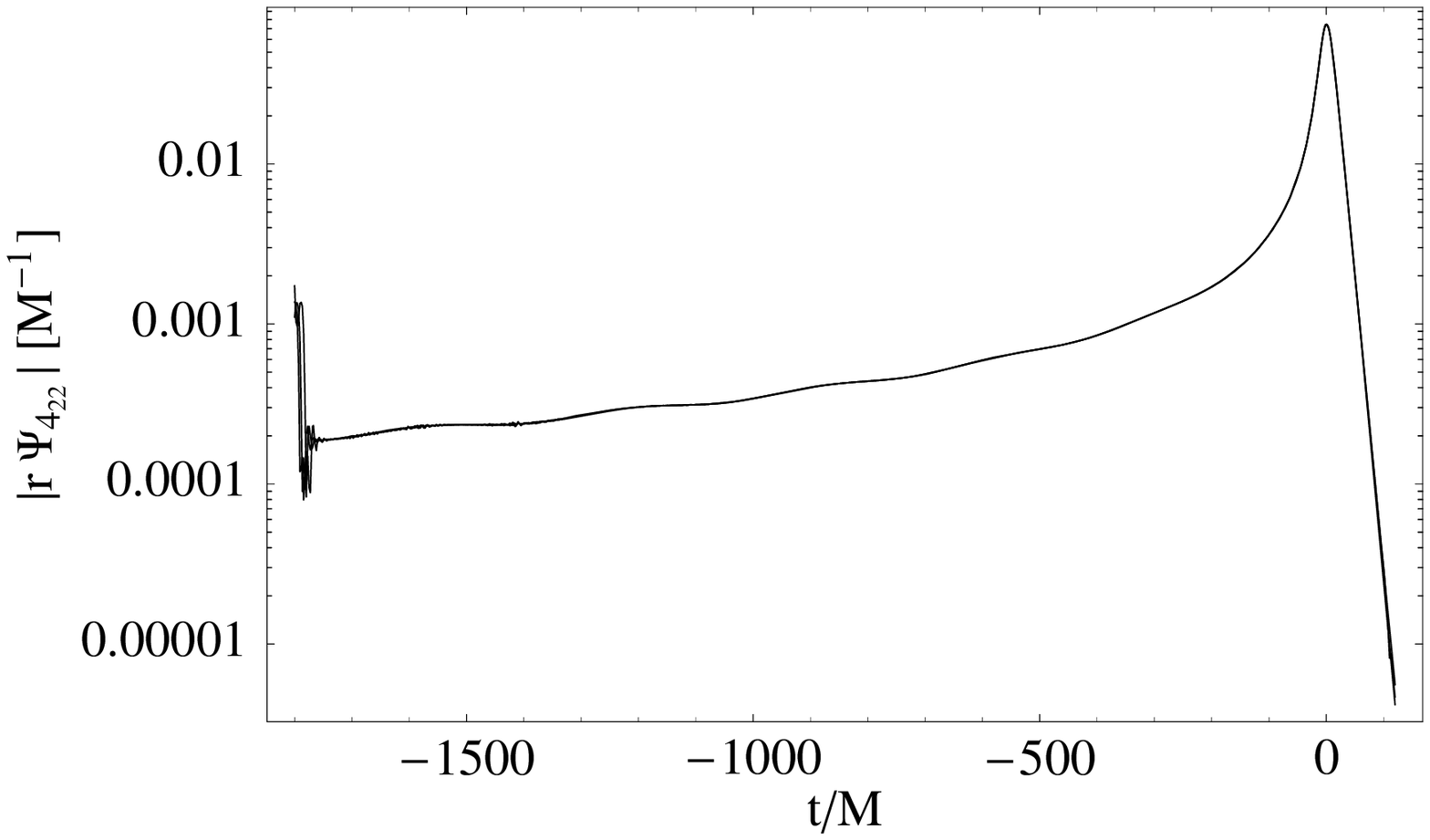}
\includegraphics[height=4cm]{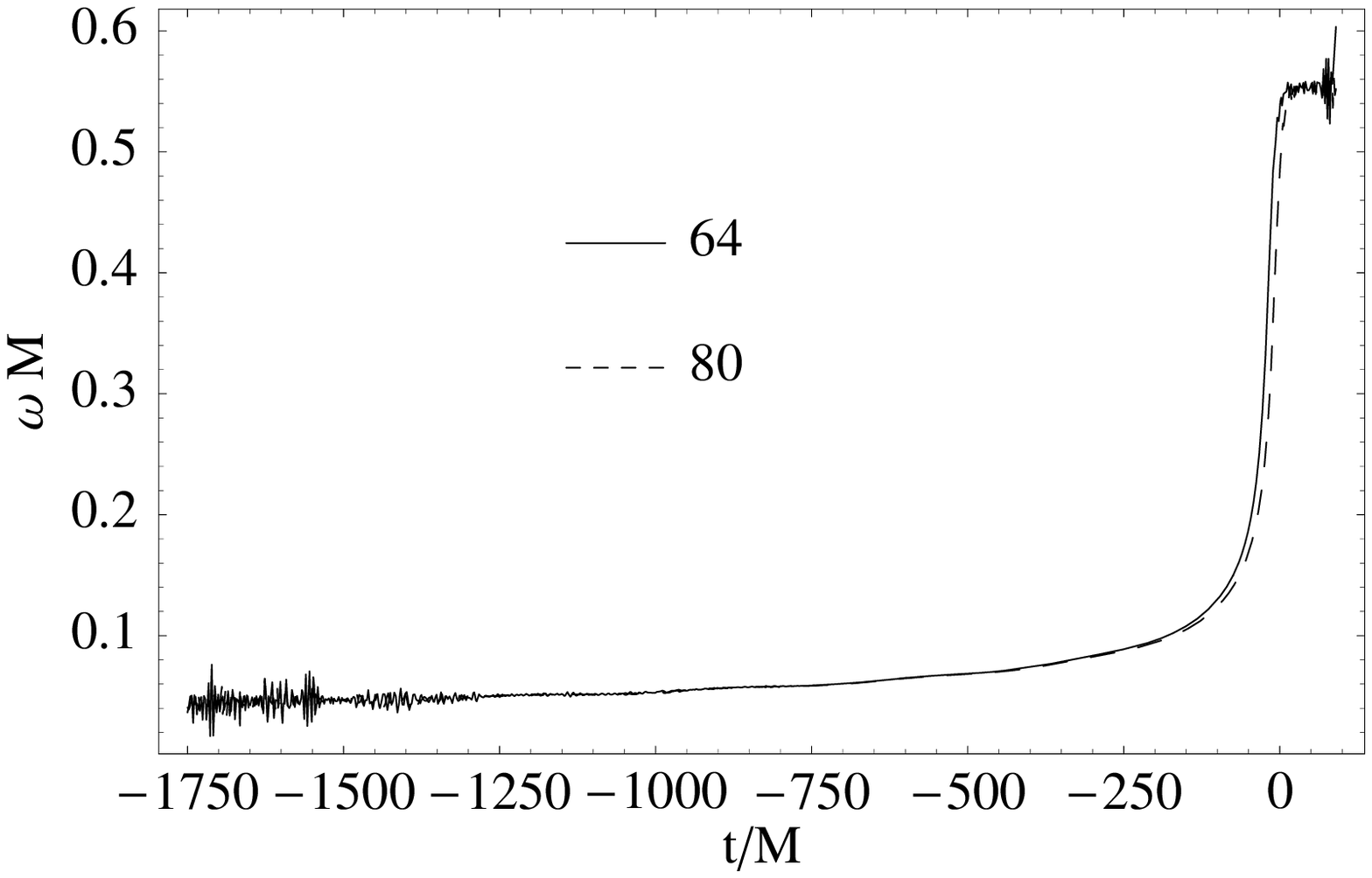}
\caption{The $l=2,m=2$ mode of the wave signal is split into the absolute
  value of $\Psi_{4,22}$ (left panel) and the wave frequency $\omega$ (right
  panel). Both panels show the simulations $\{64,72,80\}$, aligned in time to
  coincide at the peak of  $\vert \Psi_4 \vert$. The curves are clipped at
  early times, where they are very noisy due to the smallness of the signal
  and finite differencing error in computing the wave frequency from the
  phase.} 
\label{fig:wave_abs_freq}
\end{figure}
\begin{figure}[t]
\centering
\includegraphics[height=3.9cm]{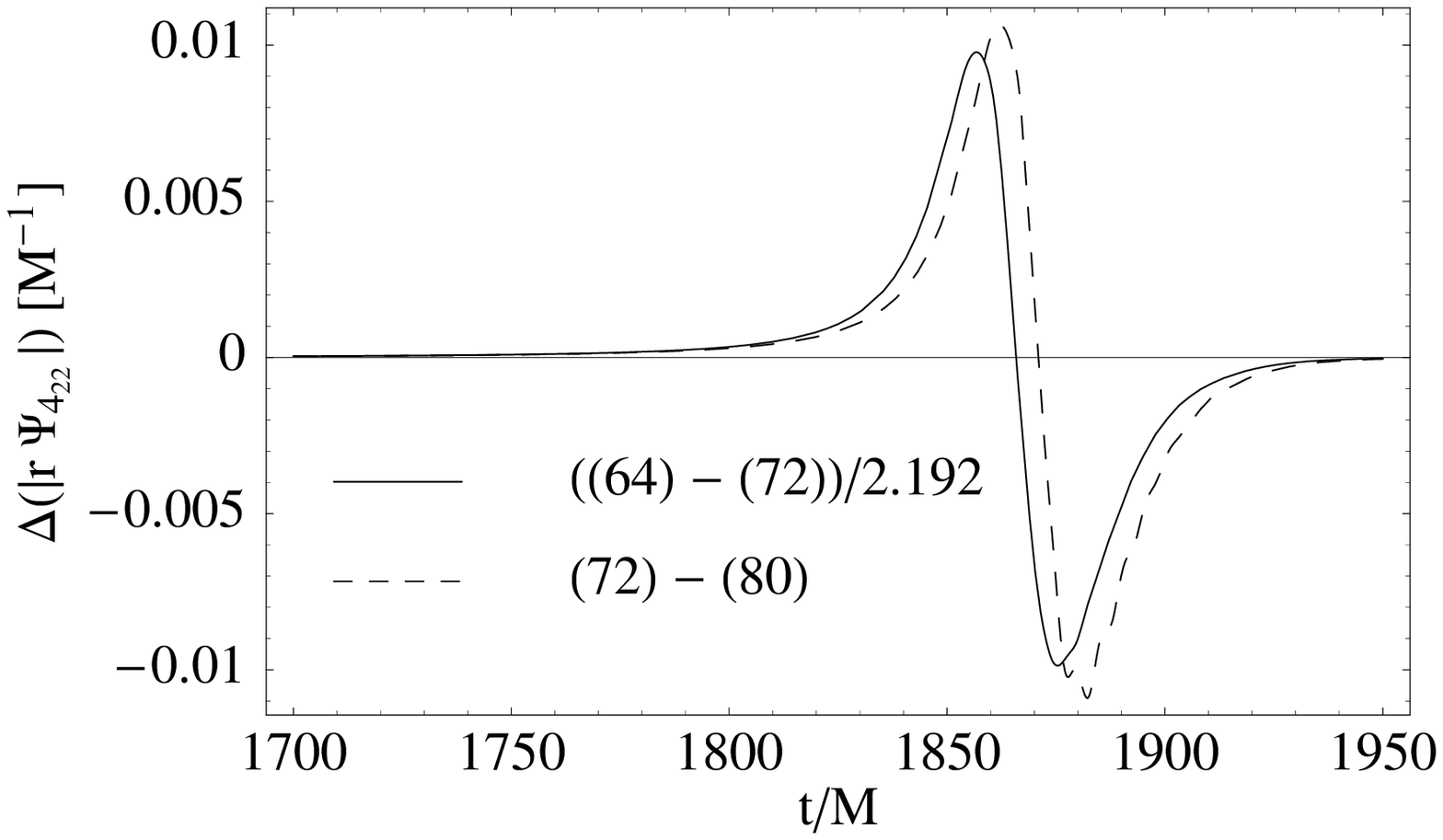}
\includegraphics[height=3.9cm]{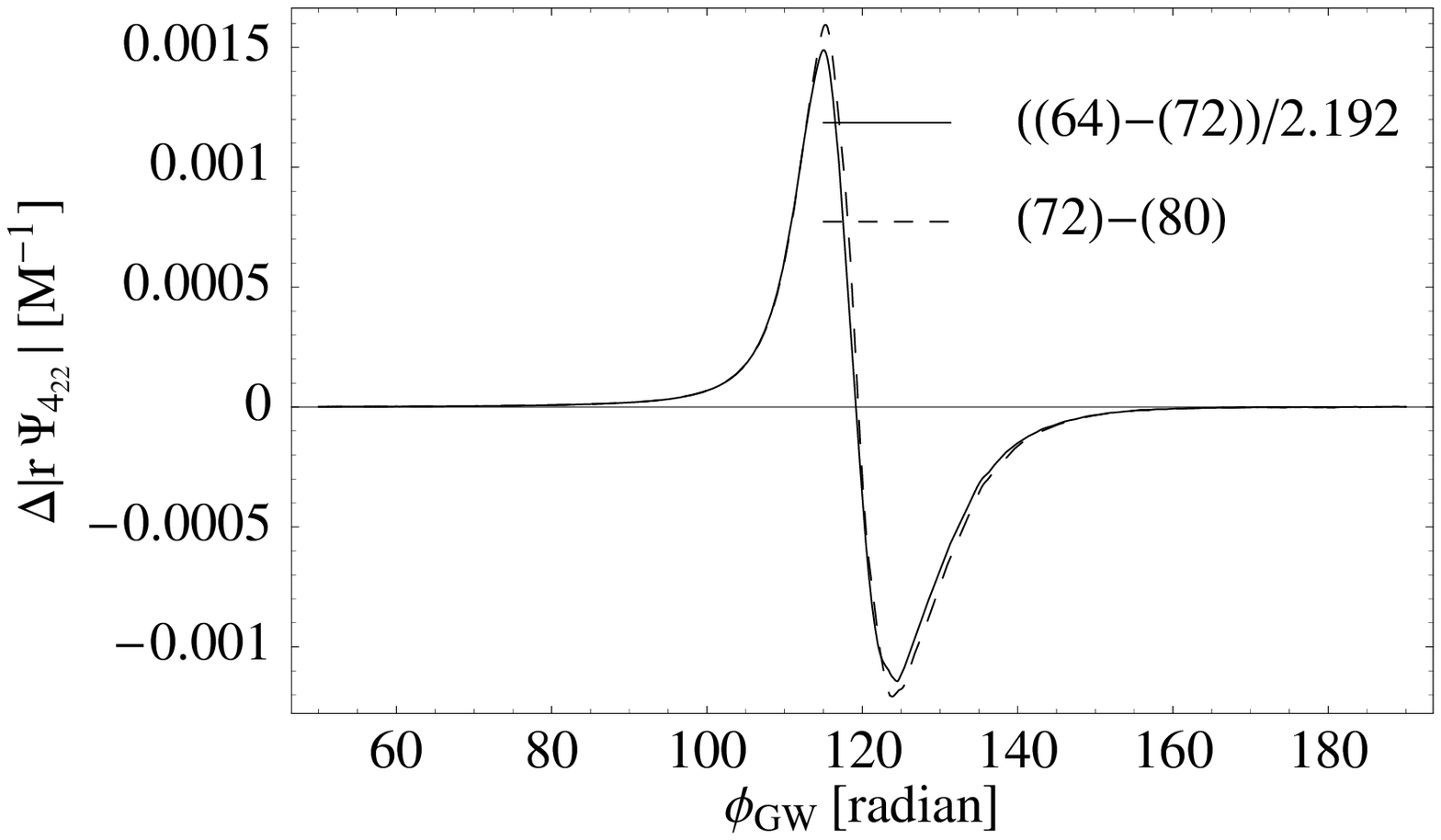}
\caption{Convergence plot for the wave amplitude $\vert {\psi_4}_{22}\vert$ in the $l=2,m=2$
mode.
Both panels show the difference between the $72$ and $80$ runs and
differences between the $64$ and $72$ runs rescaled for sixth-order convergence.
In the left panel data at different resolutions are compared at the same coordinate time,
which leads to a seeming loss of convergence near the radiation peak, which is due to the 
relatively large phase error. In the right panel
the data are compared at the same value of the gravitational wave phase, which
restores clean
sixth order convergence. 
}
\label{fig:wave_conv_amp}
\end{figure}
\begin{figure}[t]
\centering
\includegraphics[height=4cm]{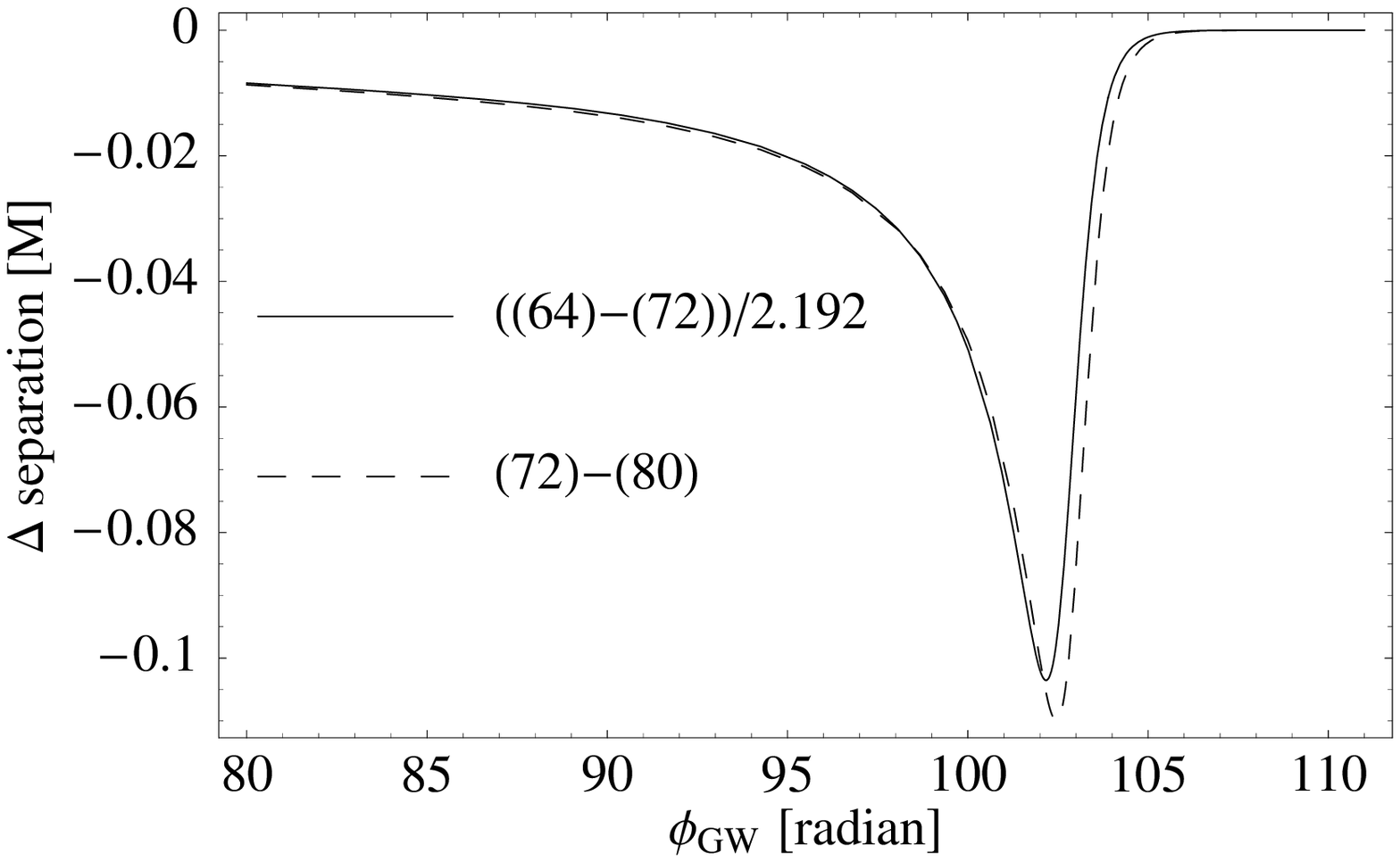}
\includegraphics[height=4cm]{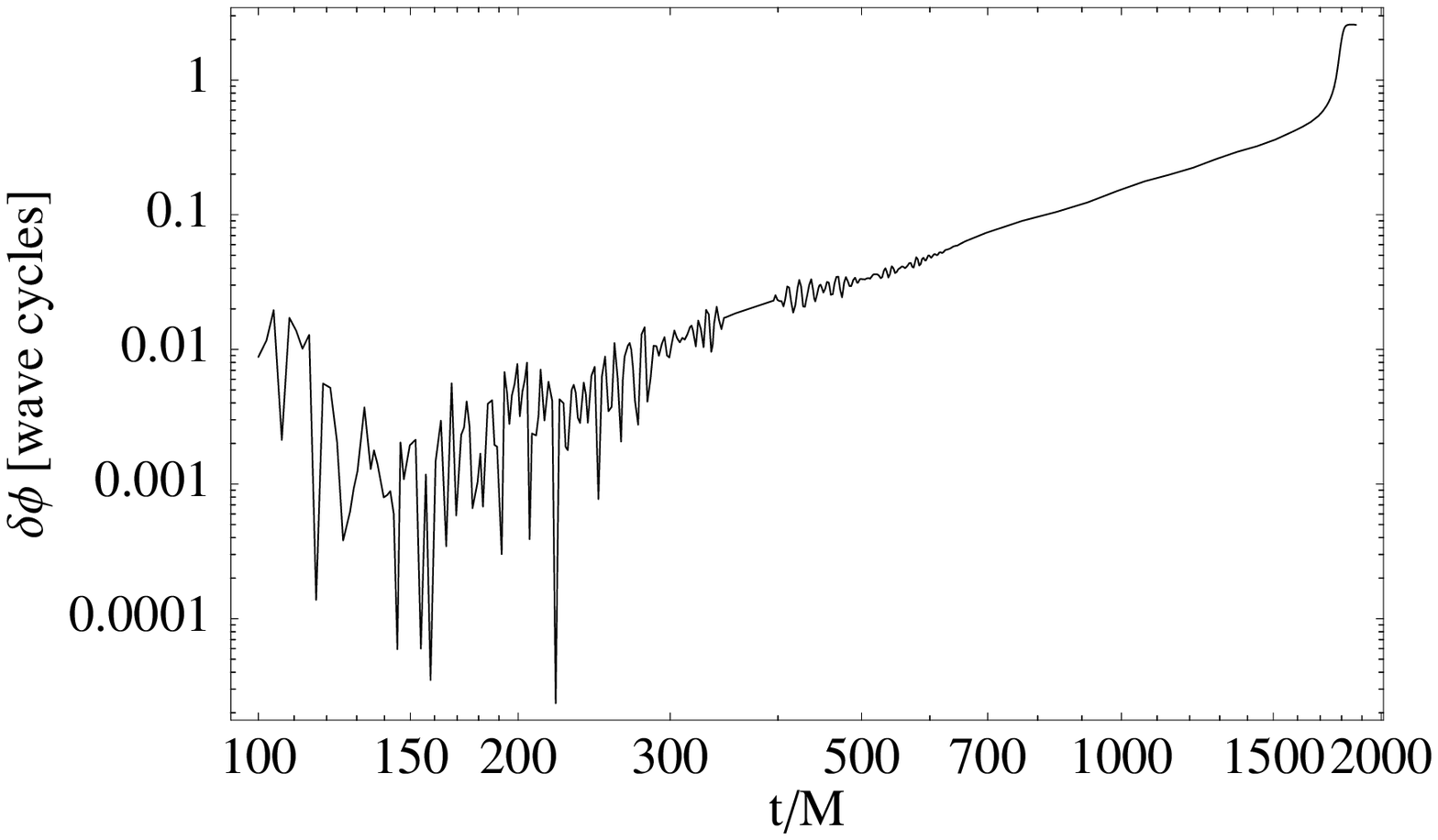}
\caption{Left panel: Reparametrisation of the black hole distance by gravitational wave phase yields
clean sixth order convergence until the time of merger (which occurs roughly at the peak of the error) --
the difference between the $72$ and $80$ runs and
differences between the $64$ and $72$ runs rescaled for sixth-order convergence lie essentially on top of
each other.
  Right panel: The difference in wave
  phase between the 4th and 6th order versions of the lowest resolution
  $48$ configuration shows the fourth-order algorithm ``falling behind''. }
\label{fig:wave_conv_freq}
\end{figure}

All runs are carried out with the symmetry $(x,y,z)
\rightarrow (-x,-y,z)$ and $(x,y,z) \rightarrow (x,y,-z)$, reducing the
computational cost by a factor of four. The Courant factor $C=\Delta t/h_l$ is
kept constant, and is set to $C=1/2$ for the inner grids, while for the outer
grids at levels 0--4 the time step  is kept constant at the value of level 3,
following our previous work \cite{Bruegmann:2006at}. 
All  runs presented here use six AMR buffer points, the same number
that we have used for our fourth-order accurate code~\cite{Bruegmann:2006at}. 
We stress that this is less than required to isolate the fine
level ``half'' timestep from time interpolation errors at the mesh-refinement
boundary, and in particular also less than required for a fully sixth-order
scheme following the approach of \cite{Lehner:2005vc}. 
The grid setups we have used for our simulations are displayed in
table~\ref{tab:orbit_grids_BAM} 
(using the naming convention introduced in~\cite{Bruegmann:2006at}).
All the runs listed here have been performed on the Kepler cluster at the
University of Jena (using Intel dual Woodcrest 
CPUs running at 2.66 GHz and an Infiniband interconnect), 
additional runs have been performed at LRZ Munich and the Doppler cluster  at
the University of Jena. We will denote the individual simulations by the
inner-box size, i.e., 48, 56, 64, 72 or 80, as indicated in bold in
table~\ref{tab:orbit_grids_BAM}.

Our initial data are chosen as follows: the initial coordinate separation of the 
punctures is chosen as $D=12M$, the horizon mass for each individual hole
is chosen as $m_i = 0.5M$, which corresponds to puncture mass parameters of $m_p = 0.488M$.
The initial momenta are obtained as $p = \pm 0.0850M$ from a 3PN-accurate
quasicircularity-condition as in~\cite{Bruegmann:2006at}. 

Our algorithm for gravitational wave extraction in terms of the Newman-Penrose scalar $\Psi_4$
has been described in~\cite{Bruegmann:2006at}. It is useful to write the signal in terms
of a time-dependent amplitude and phase as $$\Psi_4 = A(t) e^{i\varphi(t)},$$
and define the gravitational wave frequency as $\omega=\dot \varphi$.

The coordinate tracks of the puncture locations are shown in figure~\
\ref{fig:tracks_strain} for the simulations for which we have obtained
sixth-order convergence in the phase, $\{64,72,80\}$. Only during the last two
orbits are differences between the three runs distinguishable. The orbital
tracks show roughly 9 orbits before merger, and for the gravitational 
wave we obtain roughly 26 cycles before the ringdown signal becomes too noisy at $t
\approx 1960 M$. The right panel of figure~\ref{fig:tracks_strain} shows the
real part of the $l=2,m=2$ mode of the rescaled strain $r \, h_{22}$ ( 
$\Psi_4 = \ddot h$, compare \cite{Ajith:2007qp}).

Figure~\ref{fig:distance_phase} shows the coordinate distance and
gravitational wave phase for the black holes for the $\{64,80\}$ 
simulations (the $72$ simulation would not be distinguishable from the $80$
run on the scale shown here). We find that lower resolutions merge earlier,
and this is systematic for all the runs we have performed. 
For the $\{48,64,72,80\}$ runs we obtain ``merger times'' of $t=(1746.8,
1790.7, 1797.5, 1800.8)M$, and the maximum of the radiation power is reached
at times $t=(1818.5,1862.2,1869.0,1872.4)M$. ``Merger time'' here is
understood only as a rough indicator of the time of merger, which is used for
convergence tests, and is defined as the time when the coordinate distance of
the punctures drops below $0.5 M$. For the $\{64,72,80\}$ runs the merger times
show convergence at order $5.55$, the radiation peak times show convergence at
order $5.4$. Richardson extrapolation with convergence order $5.5$ yields an
error estimate of $\approx 4 M$ for both times. Note that oscillations, which
are probably mainly due to eccentricity, can clearly be seen in the black hole
distance, and also in the wave amplitude shown in
figure~\ref{fig:wave_abs_freq}. A method to reduce eccentricity will be
discussed in \cite{Husa:2007ec}. 

We have obtained roughly sixth order convergence for the gravitational wave phase
between about $t=1000 M$ and $t=1800 M$, as shown in  figure~\ref{fig:wave_conv}.
At earlier times the convergence factor becomes very noisy due to the
smallness of the signal. Shortly before the merger the convergence factor
``glitches'' to a value of roughly $7$. This problem can also clearly be 
seen in the convergence of the radiation frequency and amplitude as shown in  
figures~\ref{fig:wave_conv_freq} and \ref{fig:wave_conv_amp}. 
This ``glitch'' appears when the frequency and phase increase very sharply,
and small phase errors have a large effect. 
The logarithmic scaling version of the convergence plot in figure~\ref{fig:wave_conv}
shows a slow and rather clean exponential growth for the phase error at
intermediate times, a nonlinear fit for $ 300M \leq t \leq 1400M$  
yields $\delta \varphi = 0.0117 \exp(0.003 t/M)$ for the phase error. This
observation provides one way to optimize numerical methods in the inspiral
phase, without evolving all the way to the merger. In the last stage of the
inspiral the phase error grows very rapidly. We have noted this previously in 
a different context in~\cite{Bruegmann:2006at}, where we have compared
different methods to provide quasicircular inspiral data. Small changes on the
order of 1\% of the initial momenta have lead to drastic changes of $\approx
40 M$ in merger time. 
In order to clarify the convergence behaviour of our code,
we have applied a new technique to check for convergence in situations where
the numerical error is dominated by phase shift: We first perform a convergence
analysis for the dependence of the gravitational (or orbital) phase on code time, and 
then perform a standard convergence test on a quantity like the
puncture separation (Fig. (\ref{fig:wave_conv_freq})) or wave amplitude (Fig. (\ref{fig:wave_conv_amp}))
regarding the functional dependence of this quantity on the phase. 

An important question aside from convergence is how the sixth order and fourth
order algorithms compare in absolute numbers. For this purpose we have re-run
the $48$ configuration with our standard fourth order algorithm. We find
that already at this 
low resolution the sixth order algorithm is superior, as shown in
figures~\ref{fig:distance_phase} and \ref{fig:wave_conv_freq}, while at higher 
resolutions the larger convergence factor increases the gain in accuracy.

\section{Conclusions}

We have described a minimal extension of the fourth-order accurate
evolution algorithm described in~\cite{Bruegmann:2006at}, where by
replacing spatial fourth-order accurate differencing operators in the
bulk of the grid by sixth-order accurate stencils, we gain drastic
improvements in accuracy for the phase in long simulations of
equal-mass inspiral.  The crucial technical point regarding the choice
of sixth-order accurate finite difference operators has been a
specific choice for the advection stencil, which is used to discretize
Lie derivative terms with respect to the shift vector in the Einstein
equations.  Using this method we have demonstrated evolutions of about
nine orbits or $1800M$ with a phase error of approximately $4 M$ in
the merger time, requiring $\approx 11100$ CPU hours on our in-house
cluster.
We emphasize that our code has several lower-order accurate
ingredients, which however do not seem to contribute significantly to
the numerical error at the resolutions we employ. 

Our emphasis here has been on boosting the current generation of
``moving puncture'' codes regarding their efficiency to analyse
physical situations that require long evolutions, such as an accurate
comparison with post-Newtonian results (see~\cite{Hannam:2007}),
rather than on numerical analysis. Some technical questions certainly
remain, such as the reduction of the numerical error at mesh
refinement boundaries, the optimization for different architectures,
and a rigorous mathematical analysis of numerical stability, e.g.,\ by
extending~\cite{Calabrese:2005ft} to the BSSN system and the
complications arising in the context of mesh-refinement.
In future work we will present applications of our algorithm to other
situations such as unequal masses and evolutions of spinning black
holes.

\ack

This work was supported in part by
DFG grant SFB/Transregio~7 ``Gravitational Wave Astronomy''.
We thank the DEISA Consortium (co-funded by the EU, FP6 project
508830), for support within the DEISA Extreme Computing Initiative
(www.deisa.org); computations were performed at LRZ Munich and the Doppler and Kepler
clusters at the Institute of Theoretical Physics of the University of Jena.

\section*{References}
\bibliographystyle{unsrt}
\bibliography{refs}
\end{document}